\documentclass[aps,pre,twocolumn,amsmath,amssymb,superscriptaddress,showpacs,longbibliography]{revtex4-1}

\usepackage{graphicx}
\usepackage{dcolumn}
\usepackage{bm}
\usepackage{mathrsfs}

\usepackage{natbib}
\usepackage[text={7in,9.5in},centering]{geometry}

\usepackage{hyperref}
\usepackage{url}

\usepackage{epsfig}
\usepackage{amsmath}
\usepackage{amssymb}
\usepackage{textcomp}

\usepackage{color}
\usepackage{float}

\usepackage{stackengine}
\usepackage{verbatim}

\begin{document}

\title{Medium-spin states of the neutron-rich nucleus $^{87}$Br}

\author {B.~M.~Nyak\'o}
\author {J.~Tim\'ar}
\email [Corresponding author:] {timar@atomki.hu}
\author {M.~Csatl\'os}
\author {Zs.~Dombr\'adi}
\author {A.~Krasznahorkay}
\author {I.~Kuti}
\author {D.~Sohler}
\author {T.~G.~Tornyi}
 \affiliation{Institute for Nuclear Research (Atomki), Pf. 51, 4001 Debrecen, Hungary}

\author {M.~Czerwi\'nski}
\author {T.~Rz\c{a}ca-Urban}
\author {W.~Urban}
\author {P.~B\c{a}czyk}
 \affiliation{Faculty of Physics, University of Warsaw, ul. Pasteura 5, PL-02-093 Warsaw, Poland}
 
\author {L.~Atanasova}
 \affiliation{Department of Med. Physics and Biophysics, Medical University - Sofia, 1431, Sofia, Bulgaria}
 
\author {D.~L.~Balabanski}
 \affiliation{ELI-NP, Horia Hulubei National Institute for R$\&$D in Physics and Nuclear Engineering IFIN-HH, 077125 Bucharest-Magurele, Romania}

\author {K.~Sieja}
 \affiliation{Universit\'e de Strasbourg, IPHC, Strasbourg, France; CNRS, UMR7178, 67037 Strasbourg, France}

\author {A.~Blanc}
\author {M.~Jentschel}
\author {U.~K\"oster}
\author {P.~Mutti}
\author {T.~Soldner}
 \affiliation{Institut Laue-Langevin, 71 avenue des Martyrs, 38042 Grenoble Cedex 9, France}

\author {G.~de~France}
 \affiliation{GANIL, CEA/DSM-CNRS/IN2P3, Bd Henri Becquerel, BP 55027, F-14076 Caen Cedex 5, France}

\author {G.~S.~Simpson}
 \affiliation{LPSC, Universit\'e Joseph Fourier Grenoble 1, CNRS/IN2P3, Institut National Polytechnique de Grenoble, 
F-38026 Grenoble Cedex, France}
 
\author {C.~A.~Ur}
 \affiliation{Extreme Light Infrastructure-Nuclear Physics (ELI-NP)/IFIN-HH, 077125
Bucharest-Magurele, Romania}
 
\date{\today}

\begin{abstract} 

Medium-spin excited states of the neutron-rich nucleus $^{87}$Br were observed and studied for the first time. 
They were populated in fission of $^{235}$U induced by the cold-neutron beam of the PF1B facility of the 
Institut Laue-Langevin, Grenoble. The measurement of $\gamma$ radiation following fission has been performed 
using the EXILL array of Ge detectors. The observed level scheme was compared with results of large valence 
space shell model calculations. The medium-spin level scheme consists of three band-like structures, which 
can be understood as bands built on the ${\pi}f_{5/2}$, ${\pi}(p_{3/2}+f_{5/2})$ and ${\pi}g_{9/2}$ configurations. 
The behavior of the observed ${\pi}g_{9/2}$ band at high spins shows a considerable deviation from the shell model 
predictions. This deviation in this band is probably the result of an increased collectivity, which can be 
understood assuming that the ${\pi}g_{9/2}$ high-$\it j$ proton polarizes the core.
\end{abstract}


\maketitle 

\section{introduction}

Study of exotic, neutron-rich nuclei is in the forefront of the contemporary nuclear-structure research. Such
investigations resulted in the observation of interesting new phenomena, like quenching of the known shell 
closures and formation of new ones. Thus, it is of special interest to study neutron-rich nuclei near the 
shell closures. Such an interesting region is that of the nuclei near the doubly magic exotic $^{78}$Ni.
Besides the strength of the shell closures, the shell-model single-particle energies and residual interactions 
between the nucleons are also changing in these regions.
Moreover, deformed intruder configurations could appear which are predicted to be pushed down in energy due 
to neutron-proton correlations with enhanced quadrupole collectivity. Indeed, the very recent results for 
$^{78}$Ni indicate the breakdown of the neutron magic number 50 and proton magic number 28 beyond this 
stronghold, caused by a competing deformed structure \cite{taniuchi}. Above Z=28 the first spectroscopy 
of low-lying levels in $^{82,84}$Zn \cite{shand} at N=52,54 suggests that magicity is strictly confined 
to N=50 in $^{80}$Zn with an onset of deformation developing towards heavier Zn isotopes. 
Excited states of odd-mass nuclei provide important 
information on the single-particle energies. However, very few medium-spin excited states are known for the 
odd-mass odd-proton nuclei with neutron number larger than 50 and proton number less than 37 near $^{78}$Ni.
In general, information on excited medium-spin states in this region is very scarce, due to the difficulties 
in finding nuclear reactions in which these states could be populated with sufficient statistics. 

In this work we aimed at studying the medium-spin states of the nucleus $^{87}$Br, which can be populated 
in the cold-neutron induced fission of $^{235}$U with relatively high yields. Using this reaction we can 
get closer to the doubly magic $^{78}$Ni. Preliminary results from our study on the excited states of the 
odd-proton $^{87}$Br have been reported in a conference proceedings \cite{nyako}. Several levels and tentative 
spin-parities of these results have been confirmed by a new study on the low-spin states of $^{87}$Br from 
$\beta$ decay of $^{87}$Se, published during the preparation of this paper \cite{87Br}. The ground-state 
spin-parity has been assigned as 5/2$^-$ in agreement with the tentative assignment of Ref.~\cite{porquet1}. 
This assignment contradicts the earliest 3/2$^-$ assignment based on systematics, and raises 
the possibility of a deformed shape in this nucleus \cite{astier1}  providing a particular motivation 
to study also the deformation in $^{87}$Br. Recently, several studies have been reported on the low- and 
medium-spin states of some neighboring even-even, odd-odd and odd-neutron nuclei \cite{1, 6, 3, 4, 5, 2}.

\section{Experimental methods}
\label{expmet}

The experiment was performed using the PF1B cold-neutron beam of the Institut Laue-Langevin (ILL) in Grenoble \cite{ILL}.
The neutron beam was shaped into a 12 mm diameter pencil beam with a thermal equivalence flux of 10$^8$/(s cm$^2$).
In two parts of the beam time a 0.525 mg/cm$^2$ and a 0.675 mg/cm$^2$ thick $^{235}$U
target (both enriched to 99.7\%) were used, the former sandwiched between 15
$\mu$m thick Zr backings and the second between 25 $\mu$m thick Be backings,
 respectively, for rapid stopping of fission fragments. This enabled an almost 
Doppler-shift free measurement of the emitted $\gamma$ rays, which were detected by the EXILL detector array \cite{jentschel}.  
It consisted of eight Compton-suppressed EXOGAM Clover detectors \cite{simpson}, six Compton-suppressed GASP 
detectors \cite{bazzacco} and two Clover detectors of the Lohengrin spectrometer \cite{simpson2}.  
The distance between the faces of detectors and target was about 15 cm. The data were collected 
using a digital acquisition system with a 100 MHz clock in a triggerless mode. Altogether 15 terabytes of data were
collected over the 21 day period of the experiment. During the offline analysis the triggerless events, each consisting of 
an energy signal and the time of its registration, were arranged into coincidence events within various time windows 
(from 200 to 2400 ns) and sorted into 2D and 3D histograms.

In order to calibrate the $\gamma$-ray energies in the coincidence matrices, we used inner calibration lines of
known transitions strongly produced in the fission process. The used calibration lines were the 199.326(6), 330.88(9),
393.7(1), 431.3(1) and 509.3(1) keV transitions from $^{144}$Ba \cite{144Ba}; the 588.825(18) keV transition from $^{138}$Xe \cite{138Xe};
the 815.0(1) and 977.8(1) keV transitions from $^{96}$Sr \cite{96Sr}; the 1133.7(3) and 2247.8(2) keV transitions from $^{135}$I \cite{135I};
as well as the 1279.0(1) and 2865.6(2) keV transitions from $^{134}$Te \cite{134Te}. The uncertainty of the calibration is
0.1 keV below 1200 keV $\gamma$-ray energy and 0.3 keV above it.

Information to assist in spin-parity assignment of the newly assigned levels could be inferred from the measured 
angular correlation relations of the subsequent transitions in the $\gamma$-decay cascades. The eight EXOGAM clover 
detectors were mounted in the EXILL spectrometer in one plane perpendicular to the beam direction in an octagonal 
geometry. The 28 detector pairs provided three different relative angles: 0$^{\circ}$, 45$^{\circ}$ and 90$^{\circ}$ 
taking into account the symmetries. The way of analyzing of angular correlations in $^{87}$Br is described in detail in 
Sec. 7.2 of Ref.~\cite{jentschel}. In particular we took into account the fact that the EXILL clover detectors were operating in the add-back mode. For this reason the attenuation coefficients in the angular-correlation formula (for example see formula 3.3 in Ref.~\cite{urban1}), were carefully determined for the EXILL clover detectors, which cover larger solid angle than typical coaxial detectors. The relevant $\chi^2$-minimization procedure is illustrated in Fig. 17 of Ref.~\cite{jentschel}.

\section{Level scheme}
In order to assign new transitions to $^{87}$Br and to build the medium-spin level scheme of this nucleus, we analyzed the
3D $\gamma$-ray histograms created by applying a 200 ns coincidence time window and built in the Radware format \cite{radware}. 
The strategy we used for assigning new transitions to $^{87}$Br was based on the fact that the transitions from the complementary 
fission fragments are in prompt coincidence with each other. In our case the main complementary fission fragments for $^{87}$Br were 
the isotopes of Lanthanum. In the cold-neutron-induced fission of $^{235}$U in most of the cases two or three neutrons and no 
protons were emitted from the primary fission fragments, leading, respectively, to $^{147}$La and $^{146}$La, as secondary fission 
fragments. Fortunately, the level schemes of these nuclei are rather well known from previous fission experiments \cite{zhu, hwang}.
In the followings we consider only these most probable types of the fission process.

As a first step, we have set double gates on several strong transition pairs of $^{147}$La or of $^{146}$La. These spectra 
are expected to contain three types of $\gamma$ rays: (a) $\gamma$ rays belonging to $^{147}$La or to $^{146}$La, respectively, 
(b) $\gamma$ rays belonging to the complementary fission fragments of $^{147}$La or $^{146}$La, which are the $^{86,87}$Br 
nuclei or the $^{87,88}$Br nuclei, respectively, and (c) $\gamma$ rays belonging to other nuclei and being in coincidence with transitions having similar energies as the used gate pair. 

The first type of $\gamma$ rays are known from previous studies \cite{zhu, hwang}. The second type of $\gamma$ rays are 
expected to appear in all the gated spectra, while those of the third type are expected to appear only in one of them. 
Four large-intensity transitions have been seen to appear in most of the coincidence spectra with gate pairs from both the 
$^{147}$La and the $^{146}$La nuclei; the 619, 663, 762 and 875 keV transitions. Thus, these transitions are expected to belong 
to the $^{87}$Br nucleus, which is the common complementary fission fragment of the two nuclei. This assignment is in agreement 
with the fact that none of these transitions are known to belong to the neighboring $^{86,88}$Br nuclei, for which the medium-spin states are better known. 

In the second step we set one of the double gates 
on one of these candidate transitions while the other gate on a strong transition of the complementary nucleus. Fig. 1
shows examples of spectra obtained in such a way. In these spectra we observed other new transitions which are in 
coincidence with the candidate $^{87}$Br transitions but do not belong to the complementary nucleus. Thus, these 
transitions probably belong also to $^{87}$Br. These transitions were very weak, or even not seen due to the background,
in the first-step spectra.

\begin{figure}[h!]
\includegraphics[width=70mm,angle=-90,bb=45 50 600 700]{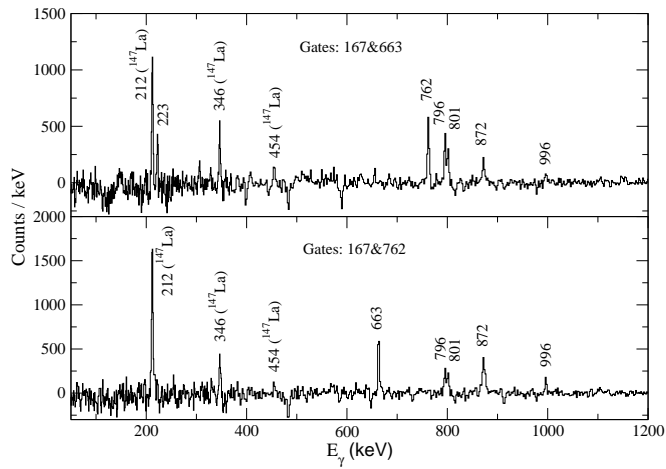}
\caption{Double-gated $\gamma\gamma\gamma$-coincidence spectra with one gate set on the strong 167 keV $^{147}$La 
transition and the
other gate set on one of the $^{87}$Br candidate transitions.}
\label{spec}
\end{figure}

Based on the coincidence relationships between these newly observed transitions, we could build a level scheme, plotted in Fig. 2.

\begin{figure}[ht]
\includegraphics[width=95mm,bb=70 100 610 720]{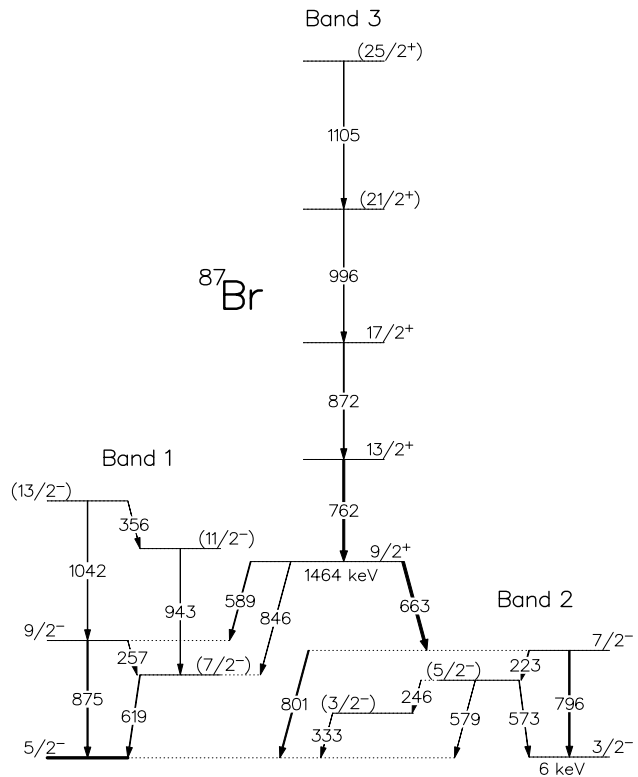}
\caption{Partial level scheme of $^{87}$Br, seen in prompt coincidence with $^{146}$La and $^{147}$La. The energies are 
given in keV, the width of the transitions are proportional with their relative intensities.}
\label{lslev}
\end{figure}
This level structure does not correspond to any known level structure belonging to the populated fission fragments. Furthermore,
when setting double gates on the strong transitions of this structure, we systematically see the strong transitions of the
$^{146}$La and $^{147}$La nuclei in the spectra, as it is shown in Fig. 3. 

\begin{figure}[h!]
\includegraphics[width=95mm,bb=40 50 600 720]{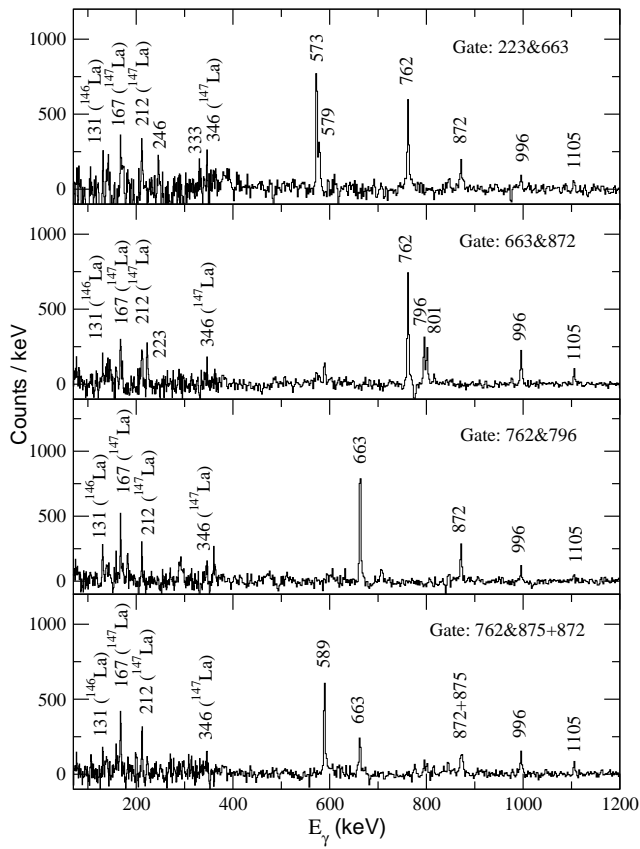}
\caption{Example double-gated $\gamma\gamma\gamma$-coincidence spectra confirming the placements of the newly observed $^{87}$Br
transitions in the level scheme.}
\label{poslev}
\end{figure}

Based on these facts, we assign the obtained level scheme to $^{87}$Br. In this partial level scheme the levels correspond to 
the states populated from prompt fission. The major part of this partial level scheme has been reported in a conference
proceedings~\cite{nyako}. In this level scheme we extended Bands 1 and 3 by one level each, and adopted the 333 keV level from Ref.~\cite{87Br} compared to the level scheme in Ref.~\cite{nyako}. The levels up to the 1464 keV one and the transitions depopulating them have been confirmed by Ref.~\cite{87Br}. The obtained level energies, spin-parities, $\gamma$-ray energies and relative intensities, as well as the $\gamma$-ray branching ratios are given in Table~I. 

\begin{table}[h!]
\caption{\label{tab:tab1}Level energies, spin-parities, $\gamma$-ray energies and relative intensities, as well as $\gamma$-ray
branching ratios corresponding to the observed medium-spin level scheme of $^{87}$Br. The relative intensities are normalized to
that of the strong 662.9 keV transition. nb denotes a level not belonging to any band.}
\begin{ruledtabular}
\begin{tabular}{ccccccc}
$E_i$ & $ I^{\pi}_i $ & Band& $E_f$ & $ E_{\gamma} $&$ I_{\gamma} $ & BR\\
\hline \\[-7pt]
0.0 & 5/2$^-$ & 1\\
6.0(3) & 3/2$^-$ & 2\\
332.9(4) & (3/2$^-$) & nb & 0.0 & 333.1(6) & $>$2\\
578.8(2) & (5/2$^-$) & 2 & 0.0 & 578.7(3) & $>$7 & 42(9)\\
& & & 6.0 & 572.8(3) & $>$17 & 100(9)\\
& & & 332.9 & 246.0(5) & $>$2 & 12(4)\\
618.7(3)&(7/2$^-$)&1&0.0&618.9(4)& $>$26\\
801.5(2)&7/2$^-$&2&0.0&801.4(3)&46(5)\\
& & & 6.0&795.5(3)&54(5)\\
& & & 578.8&222.7(3)&26(4)\\
875.3(3)&9/2$^-$&1&0.0&875.2(5)& $>$45&100(17)\\
& & & 618.7&257.0(7)& $>$5&12(3)\\
1464.4(3)&9/2$^+$&3&618.7&845.7(6)&15(3)\\
& & & 801.5&662.9(2)&100(8)\\
& & & 875.3&589.1(3)&31(3)\\
1561.4(5)&(11/2$^-$)&1&618.7&942.7(5)&$>$4\\
1917.4(6)&(13/2$^-$)&1&875.3&1042.0(6)&19(3)\\
& & & 1561.4&356.0(7)&4(2)\\
2226.3(3)&13/2$^+$&3&1464.4&761.9(2)&77(6)\\
3098.0(4)&17/2$^+$&3&2226.3&871.7(2)&31(4)\\
4094.0(5)&(21/2$^+$)&3&3098.0&996.0(3)&12(2)\\
5199.2(6)&(25/2$^+$)&3&4094.0&1105.2(3)&7(2)\\
\end{tabular}
\end{ruledtabular}
\end{table}

In Table II we present results of angular-correlation analysis performed to help spin assignments of some of the excited levels in $^{87}$Br. Figure 4 illustrates such analysis for the 589 keV -- 875 keV cascade in $^{87}$Br. The upper panel shows the experimental A$_2$/A$_0$ and A$_4$/A$_0$ values with their uncertainties (blue box) compared against theoretical (A$_2$/A$_0$ ; A$_4$/A$_0$) points calculated for the mixing-ratios, $\delta$  ranging from minus infinity to plus infinity (Inf) represented by the ellipse (red). The green cross shows the $\delta$ value of -0.456 determined by the $\chi$$^2$ analysis illustrated in the lower panel. The calculation was done for the 9/2 $\rightarrow$ 9/2 $\rightarrow$ 5/2 spin hypothesis in the cascade. The theoretical (A$_2$/A$_0$ ; A$_4$/A$_0$) points for a stretched dipole or a stretched quadrupole 589 keV transition are outside the experimental region. For example, in the pure stretched quadrupole case the expected A$_2$/A$_0$ and A$_4$/A$_0$ values are 0.102 and 0.009, respectively.

The observed level scheme forms three band-like structures labeled as Band 1, Band 2 
and Band 3 in Fig. 2. We note that as Band 2 contains only three levels, these levels can also be non-band single-particle 
excitations. The bandheads of Band 1
and Band 2 are very close to each other in energy. The difference between them is only 6 keV. The bandhead of Band 1 is the
ground state of $^{87}$Br. Two of the bands, Band 1 and Band 2, show the feature of strongly coupled $\Delta$I=1 rotational
bands, while the third one, Band 3, looks like a weakly coupled or decoupled $\Delta$I=2 band. 

\begin{table}[h!]
\caption{\label{tab:tab2} Normalized experimental angular correlation coefficients for selected $E_{\gamma1} - E_{\gamma2}$ 
${\gamma}$ cascades in $^{87}$Br.}
\begin{ruledtabular}
\begin{tabular}{cccr}
E$_{\gamma1} - E_{\gamma2}$ & A$_2$/A$_0$ & A$_4$/A$_0$ & Spin sequence\\
\hline \\[-7pt]
662.9$-$795.5 & $-$0.05(4) & 0.05(10) & 9/2$\rightarrow$7/2$\rightarrow$3/2 \\
589.1$-$875.2 & 0.24(2) & 0.08(4) & 9/2$\rightarrow$9/2$\rightarrow$5/2 \\
871.7$-$761.9 & 0.11(6)  & $-$0.09(13) & 17/2$\rightarrow$13/2$\rightarrow$9/2 \\
\end{tabular}
\end{ruledtabular}
\end{table}

The observed angular correlations of the two lowest transitions in Band 3, i.e. the 762 keV and 872 keV transitions, are
consistent with stretched quadrupole character for both of them. This indicates that Band 3 is an E2 band, probably corresponding
to a weakly coupled or decoupled configuration. This assumption is confirmed by the fact that no cross-over transitions have been
observed in this band. Both Band 1 and Band 2 look like $\Delta$I=1 bands with quadrupole cross-over transitions. This assumption 
is also confirmed by the angular correlations observed for the 663 keV -- 796 keV cascade in Band 2 and for the 589 keV -- 875 keV 
cascade in Band 1. The observed angular correlations in the first cascade are consistent with the stretched quadrupole character
of the 796 keV transition and stretched dipole character of the 663 keV transition. In Band 1, the observed angular correlations 
are consistent with the stretched quadrupole character of the 875 keV transition and $\Delta$I=0 dipole character of the 589 keV transition. 

\begin{figure}[h!]
\includegraphics[width=100mm,bb=5 320 320 720]{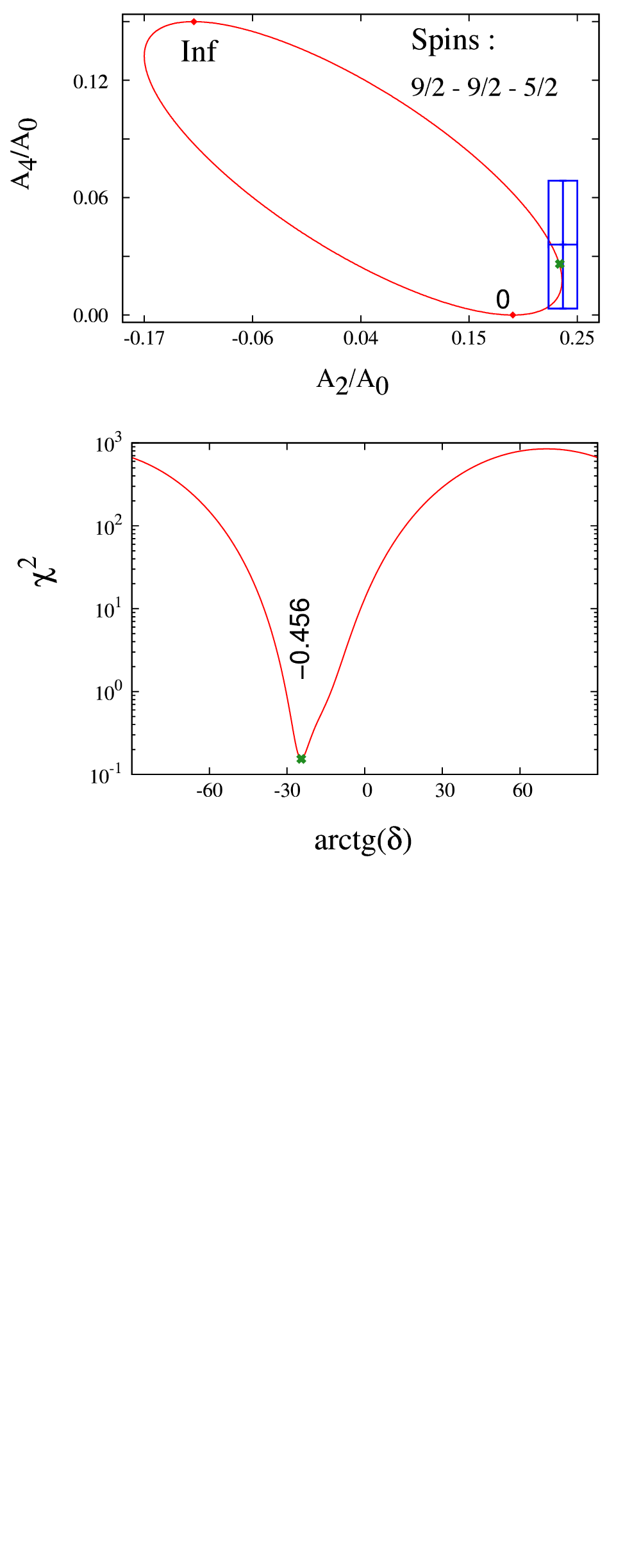}
\caption{Angular-correlation analysis for the 589 keV - 875 keV cascade in $^{87}$Br. See text for more details.}
\label{poslev}
\end{figure}

Using the multipole character of the transitions in $^{87}$Br derived in this work and the unique spin-parity assignments reported in Ref.~\cite{87Br}, together with information on the level schemes of neighboring odd-proton nuclei, we propose spin-parity values for levels in $^{87}$Br as shown in Table I and in Fig. 2. Our analysis agrees with Ref.~\cite{87Br} and provides unique spin-parity assignments for the 6 keV, 801 keV, 2226 keV and 3098 keV levels.

A $\Delta$I=2 band based on the ${\pi}g_{9/2}$ configuration is expected to appear at relatively low energy in $^{87}$Br with a 
band-head spin-parity of 9/2$^+$ based on the deformed Nilsson scheme and also on the comparison with the level schemes of the 
neighboring odd-proton nuclei. In $^{89}$Rb this band-head is at 1195 keV excitation energy \cite{89Rb}. In $^{87}$Br the band-head 
energy of Band 3 is close to this value. These arguments suggest ${\pi}g_{9/2}$ configuration to Band 3 and 9/2$^+$ spin-parity to
its band-head level. Indeed, 9/2$^+$ spin-parity has been firmly assigned to the 1464 keV level in Ref.~\cite{87Br}. The observed 
stretched quadrupole character for the two lowest transitions of this band suggests 13/2$^+$ and 17/2$^+$ spin-parities for the 
second and third levels of the band, respectively. For the two highest levels of the band only tentative spin-parities of (21/2$^+$)
and (25/2$^+$) could be assigned based on the $\Delta$I=2 band character. 
Ref.~\cite{87Br} assigned 5/2$^-$ and 9/2$^-$ spin-parities for the ground state and for the 975 keV level, respectively. These
assignments are in a good agreement with observed multipole characters of the 875 keV and the 589 keV transitions in the present
experiment. The tentative spin-parity assignments of the other levels in Band 1 have been derived from the $\Delta$I=1 band character
of this level structure.
3/2 and 7/2 spin values are assigned to the 6 keV and 801 keV levels based on the observed multipolarities of the 663 keV and 
796 keV transitions assuming that spin values increase with increasing level energy. Negative parity is assigned to these levels
because no positive-parity states are expected at such low energy. The tentative (5/2$^-$) spin-parity of the 579 keV level is
assigned assuming $\Delta$I=1 band character for Band 2. The tentative (3/2$^-$) spin-parity of the 333 keV level is taken from
Ref.~\cite{87Br}.

\section{discussion}
\label{disc}

The observed medium-spin band structures can qualitatively be understood based on the deformed, rotational picture. Indeed, 
in this picture two strongly coupled bands, based on the 5/2[303] (${\pi}f_{5/2}$) and 3/2[301] (mixed ${\pi}p_{3/2}$ and 
${\pi}f_{5/2}$) Nilsson orbitals, and one decoupled band, based on the 1/2[440] (${\pi}g_{9/2}$) Nilsson orbital, are expected 
to appear at low excitation energy with band-head spin-parities of 5/2$^-$, 3/2$^-$ and 9/2$^+$, respectively.

However, $^{87}$Br with only two neutrons above the N=50 shell closure is expected to be a shell-model nucleus. Therefore, to 
verify the suggested spin-parities and configurations, we compared them with results of the contemporary shell model calculations 
using a large valence space including the 1$f_{5/2}$, 2$p_{3/2}$, 2$p_{1/2}$, 1$g_{9/2}$ orbitals for protons and the 
2$d_{5/2}$, 3$s_{1/2}$, 1$g_{7/2}$, 2$d_{3/2}$, 1$h_{11/2}$ orbitals for neutrons, outside the $^{78}$Ni core. 

\begin{figure}[h!]
\includegraphics[width=85mm,bb=10 20 730 1020]{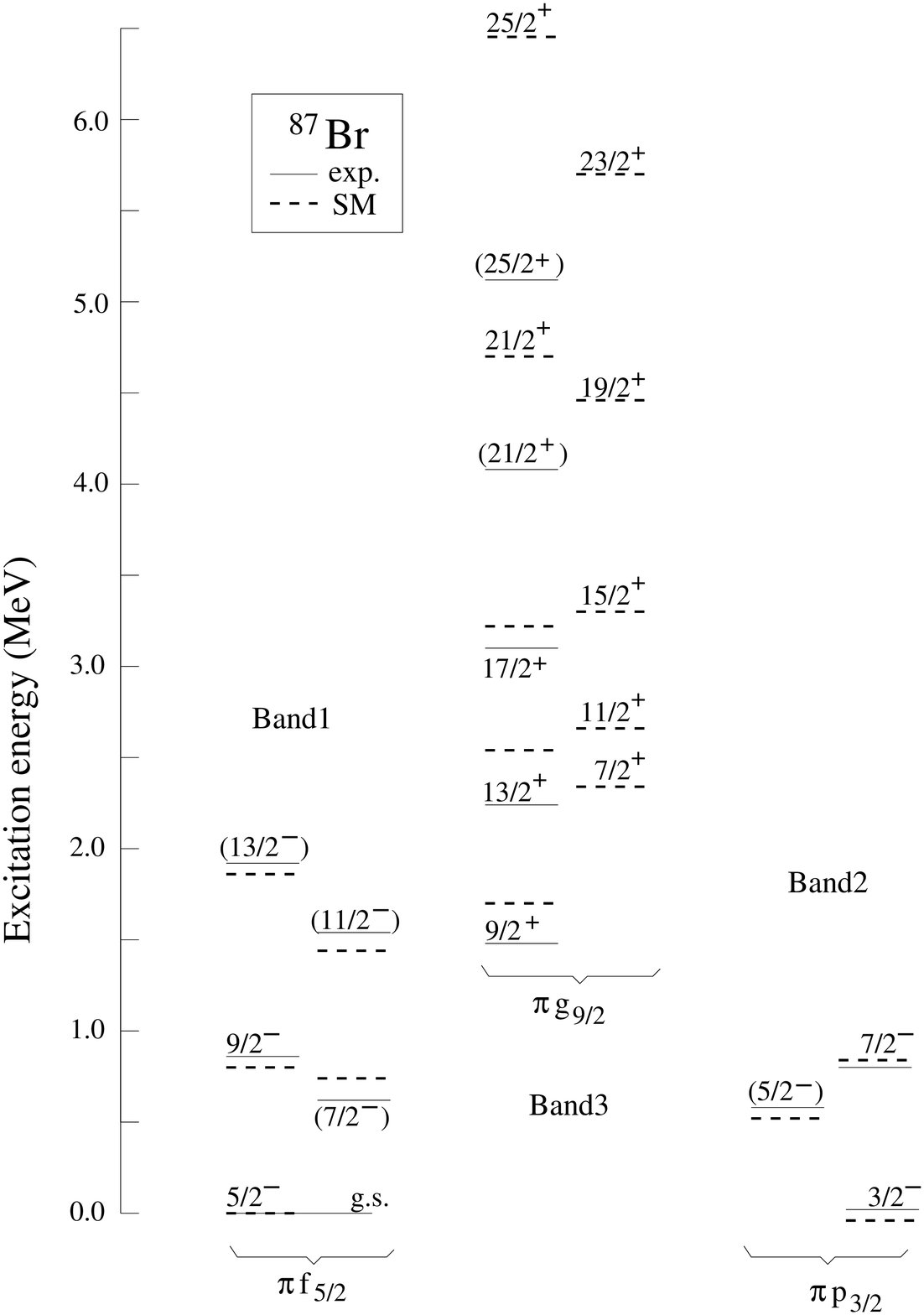}
\caption{Comparison of the observed (exp.) and calculated (SM) levels of $^{87}$Br.}
\label{exp-cal}
\end{figure}
The model and the parameters of the calculations are 
the same as used in Ref.~\cite{87Br}. The  experimental and calculated
level energies, spin-parities and single-particle configurations are plotted in 
Fig. 5.

\begin{table}[t!]
\caption{\label{tab:example1}Proton and neutron occupations of excited states in $^{87}$Br calculated using the shell model.}
\begin{ruledtabular}
\begin{tabular}{ccccccp{0.3cm}cccc} 
& \multicolumn{5}{c}{Neutrons} & & \multicolumn{4}{c}{Protons}  \\
\cline{2-6} \cline{8-11}\\[-6pt]
$I^{\pi}$ & $d_{5/2}$ & $s_{1/2}$ & $g_{7/2}$ & $d_{3/2}$ & $h_{11/2}$ & & $f_{5/2}$ & $p_{3/2}$ & $p_{1/2}$ & $g_{9/2}$ \\[2pt]
\hline\\[-10pt]\\[-8pt]
\multicolumn{11}{c}{Band1}\\
5/2$^-$ & 1.61 & 0.13 & 0.07 & 0.12 & 0.08 & &  4.48 & 1.97 & 0.33 & 0.23\\
7/2$^-$& 1.59& 0.21& 0.04& 0.11& 0.05& & 4.52& 1.96& 0.33& 0.19\\
9/2$^-$& 1.73& 0.11& 0.03& 0.09& 0.04& & 4.57& 1.90& 0.31& 0.22\\
11/2$^-$& 1.79& 0.06& 0.01& 0.11& 0.03& & 4.63& 1.87& 0.28& 0.21\\
13/2$^-$&1.85& 0.05& 0.07& 0.06& 0.02& & 4.64& 1.87& 0.28& 0.20\\[2pt]
\multicolumn{11}{c}{Band2}\\
3/2$^-$ & 1.60 & 0.15 & 0.07 & 0.10 & 0.08 & & 3.90 & 2.57 & 0.30 & 0.23 \\
5/2$^-_2$& 1.54& 0.27& 0.05& 0.09& 0.05& & 3.58& 2.80& 0.43& 0.18\\
7/2$^-_2$& 1.63& 0.20& 0.04& 0.09& 0.04& & 3.92& 2.53& 0.35& 0.20\\[2pt]
\multicolumn{11}{c}{Band3}\\
7/2$^+$& 1.56& 0.13& 0.06& 0.11& 0.14& & 3.97& 1.68& 0.30& 1.05\\
9/2$^+$& 1.57& 0.14& 0.08& 0.14& 0.07& & 4.03& 1.57& 0.28& 1.12\\
11/2$^+$& 1.73& 0.08& 0.03& 0.10& 0.06& &  4.08& 1.53& 0.30& 1.09\\
13/2$^+$& 1.56& 0.22& 0.04& 0.13& 0.04& & 3.90& 1.68& 0.31& 1.11\\
15/2$^+$& 1.75& 0.09& 0.02& 0.04& 0.10& & 4.37& 1.36& 0.22& 1.05\\
17/2$^+$& 1.78& 0.05& 0.02& 0.09& 0.07& & 4.30& 1.41& 0.22& 1.08\\
19/2$^+$& 1.22& 0.03& 0.03& 0.05& 0.67& & 4.66& 1.59& 0.24& 0.51\\
21/2$^+$& 1.76& 0.06& 0.03& 0.12& 0.03& & 3.82& 1.83& 0.29& 1.07\\
23/2$^+$& 1.74& 0.02& 0.02& 0.04& 0.18& & 4.48& 1.46& 0.14& 0.91\\
25/2$^+$& 1.31& 0.02& 0.58& 0.05& 0.03& & 3.84& 1.83& 0.26& 1.07\\
\end{tabular}
\end{ruledtabular}
\end{table}

It can be seen in the figure that the tentative experimental spin-parities 
are rather well reproduced by the calculations. The level energies are also relatively 
well reproduced for the states belonging to Band 1 and Band 2. There are, however, 
larger differences between the experimental and calculated level energies for the 
levels of Band 3, especially for its highest-energy levels. Adjusting the proton 
$g_{9/2}$ -- neutron $d_{5/2}$ interaction in the model, the band-head energy of 
Band 3 could be reproduced, however the energy spacing between the consecutive levels 
would not be changed, thus we cannot get good agreement between the experimental and 
calculated energies. It is worth mentioning that while in Bands 1 and 2 all the 
predicted levels have been observed up to the highest observed spins, in case of 
Band 3 only the $\alpha = +1/2$ signature branch is observed. This is probably due 
to the fact that the $\alpha = -1/2$ signature branch is strongly non-yrast, thus 
those levels have not been excited in the experiment. This assumption is in 
agreement with the calculations, which predict this signature branch to be shifted 
up by about 1 MeV. Table III shows the calculated proton and neutron occupations of 
the levels. They confirm the proposed single-particle configuration assignments. 

\begin{figure}[h!]
\includegraphics[width=90mm,bb=60 110 600 700]{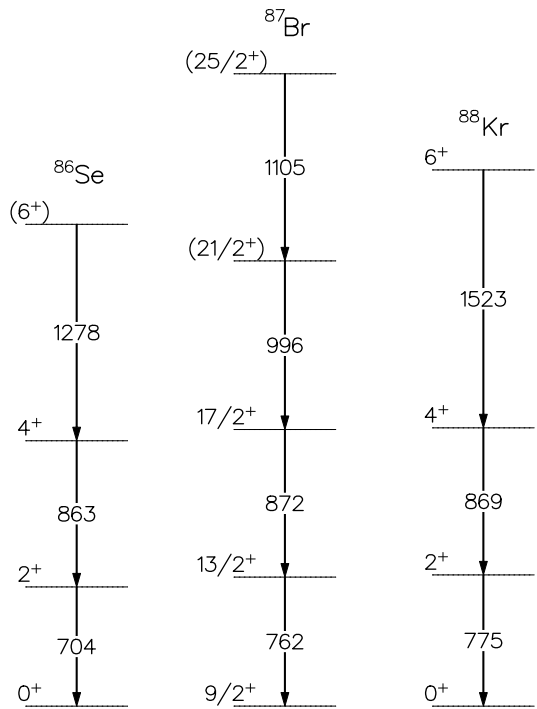}
\caption{Comparison of Band 3 in $^{87}$Br with the ground-state bands of the neighbouring N=52 even-even nuclei,  $^{86}$Se and $^{88}$Kr.}
\label{complev}
\end{figure}

The deviation between the measured and calculated level energies in Band 3 is probably due to a larger
collectivity 
than that predicted by the calculations. The larger collectivity can be understood assuming that the proton in the 
${\pi}g_{9/2}$ high-$\it j$ orbital polarizes the core. In this case, the quadrupole-quadrupole interaction mixes 
other orbitals from the next shell, e.g. 2$d_{5/2}$, with the $g_{9/2}$ one in the configuration of Band 3. Such 
orbitals are not included in the model space of the present theory, thus the calculations cannot take the effect 
of them into account. This effect of the $g_{9/2}$ proton in increasing the collectivity seems to be supported also 
by the comparison of Band 3 with the ground-state bands of the also $N=52$ $^{86}$Se and $^{88}$Kr nuclei. 
The data for the latter two bands 
were taken from Ref. \cite{1}. This comparison is shown in Fig. 6, 
where the bandhead energy of the ${\pi}g_{9/2}$ band of $^{87}$Br is shifted to match with that of the other two 
bands. It is seen that the levels of the ${\pi}g_{9/2}$ band are regularly spaced. Also, the energies of the higher-spin states 
are considerably lower than predicted by the shell model calculation, as it is seen in Fig. 5. These are characteristics for the 
increased collectivity. 
Contrarily, the levels of the $^{86}$Se and $^{88}$Kr bands, which have no ${\pi}g_{9/2}$ contribution in their configurations, are 
irregular and follow the shell model predictions as discussed in Ref.~\cite{1}. In fact, as discussed in Ref.~\cite{Litzinger} 
the ${\pi}g_{9/2}$ contribution in the configurations of these states is small but still existing.

\section{summary}
Medium-spin excited states of the neutron-rich $^{87}$Br nucleus have been studied by means of in-beam $\gamma$ 
spectroscopy of $^{235}$U(n,f) fission fragments, using the EXILL Ge detector array. The fission has been induced 
by the cold-neutron beam of the PF1B facility of the Institut Laue-Langevin, Grenoble. Medium-spin level scheme of 
this nucleus was built for the first time. The observed excited states form three band-like structures. 
Angular correlation information from the experiment, combined with systematics
of the neighbouring odd-proton nuclei enabled tentative spin-parity assignments to the new levels and configuration assignments to
the bands. They can be understood as bands built on the ${\pi}f_{5/2}$, ${\pi}(p_{3/2}+f_{5/2})$ 
and ${\pi}g_{9/2}$ configurations.
These assignments have been confirmed by the results of contemporary shell model calculations using a large valence space.
$^{87}$Br has 5/2$^-$ ground state spin-parity contrary to the odd-mass Br isotopes containing fewer neutrons,
which have 3/2$^-$ ground state spin-parity. Properties of the ${\pi}g_{9/2}$ band show an increased collectivity compared to the
other bands in $^{87}$Br and in the neighboring nuclei. This increased collectivity in this band can be understood assuming
that the ${\pi}g_{9/2}$ high-$\it j$ proton polarizes the core.

\begin{acknowledgments}
The authors thank the technical services of the ILL, LPSC, and GANIL for supporting the EXILL campaign. The EXOGAM Collaboration 
and the INFN Legnaro are acknowledged for the loan of Ge detectors.
This work was supported by the National Research, Development and Innovation Fund of Hungary, financed under the K18 funding 
scheme with project nos. K128947 and K124810, as well as by the European Regional Development Fund (Contract No. 
GINOP-2.3.3-15-2016-00034).
This work was also supported by the Polish National Science Centre under Contract No. DEC-2013/09/B/ST2/03485 and by the Bulgarian Ministry of Education and Science under the National Research Program "Young scientists and postdoctoral students". I.K. was supported by National Research, Development and Innovation Office – NKFIH, contract number PD 124717.
D.L.B. acknowledges support from the Extreme Light Infrastructure Nuclear Physics (ELI-NP) Phase II, a project co-financed by the Romanian Government and the European Union through the European Regional Development Fund, the Competitiveness Operational Programme (1/07.07.2016, COP, ID 1334).

\end{acknowledgments}

\end{document}